\documentclass[12pt]{article}

\begin{document}
\begin{titlepage}

\large
\centerline{\bf The Determination of the CKM Matrix}
\normalsize

\vskip 2.0cm
\centerline{Frederick J. Gilman$^{\dag{\,}*}$ }

\centerline{\it Department of Physics} 
\centerline{\it Carnegie Mellon University} 
\centerline{\it Pittsburgh, Pennsylvania 15213}
\vskip 4.0cm

\centerline{\bf Abstract}
\vskip 1.0cm
A general discussion of the Cabibbo-Kobayashi-Maskawa (CKM) matrix 
is given and the importance stressed of determining the matrix elements 
as an essential part of understanding CP violation in and beyond 
the Standard Model.  The status of knowledge of the matrix elements 
connecting the first and second generation quarks is reviewed.  
A perspective on determinations of the full CKM 
matrix is presented as an introduction to the separate contributions 
to the panel discussion that follows.
\vfill

\footnoterule
\noindent $^\dag$\footnotesize{\,}Electronic address: 
gilman@cmuhep2.phys.cmu.edu \\
\noindent $^*${\,}To be published in Proceedings of Beauty 2000, 
Kibbutz Maagan, Israel, September 13-18, 2000, edited by S. Erhan, 
Y. Rozen, and P. Schlein, Nucl. Inst. Meth. A, 2001.

\end{titlepage}
\newpage

\section{Introduction}

In the Standard Model with $SU(2) \times U(1)$ as 
the gauge group of electroweak interactions, both 
the quarks and leptons are assigned to be left-handed 
doublets and right-handed singlets. The quark mass 
eigenstates differ from the weak eigenstates, and
the matrix relating these bases was defined for six 
quarks and given an explicit parametrization by Kobayashi and Maskawa{\,}\cite{KobayashiMaskawa73} in 1973.  It generalizes 
the four-quark case, where the matrix is parametrized by 
a single angle, the Cabibbo angle{\,}\cite{Cabibbo63}.
 
By convention, the mixing is usually expressed in terms of a 
$3\times 3$ matrix $V$ operating on the charge $-e/3$ 
quark mass eigenstates ($d$, $s$, and $b$):
\begin{equation}
\left(\matrix{d ^{\,\prime}   \cr
                s ^{\,\prime} \cr
                b ^{\,\prime} \cr
       		}\right)
=
	\left(\matrix{
		V_{ud}&     V_{us}&     V_{ub}\cr
		V_{cd}&     V_{cs}&     V_{cb}\cr
		V_{td}&     V_{ts}&     V_{tb}\cr
       		}\right)
	\left(\matrix{
		d \cr
                s \cr
                b \cr
       		}\right) ~.
\end{equation}

     The matrix V is unitary, and this 
Cabibbo-Kobayashi-Maskawa (CKM) matrix can physically be fully 
specified by four real parameters.  These can be taken to be 
three ``rotation'' angles and one phase.  CP violation has 
a natural place and occurs if the phase is not $0^o$ or $180^o$ 
and the other angles are not $0^o$ or $90^o$, {\em i.e.}, if there 
is mixing between each pair of generations of quarks and there 
is a non-trivial phase.  

     We know experimentally that all three angles that characterize 
the CKM matrix are small but non-zero.  There is an expectation that 
the single non-trivial phase should be non-zero as well.   
(We will return at the end to the status of showing whether the 
phase is non-zero.)  If CP violation arises from the CKM matrix, 
there is both a natural scale and a special pattern for CP-violating 
effects (and for flavor-changing-neutral-current effects more generally).  
These are made manifest in the theoretical predictions for various 
decay and mixing processes for B mesons found throughout the Proceedings 
of this conference. 

    The major outstanding question with regard to CP violation is no 
longer what was raised at conferences for many years, namely 
``What is the origin of CP violation?''  We have an origin in the CKM 
matrix of the Standard Model.  It is not unreasonable that this 
accounts for most of the CP-violating effects observed to date and 
to be observed in the near future.  Rather, the question to be 
answered by experiment and theory in the coming decade is:  
``Are there CP-violating effects that do not arise from the CKM 
matrix and instead come from physics beyond the Standard Model.'' 

     Thus we expect that the situation we will soon find ourselves 
in is one where the largest measured CP-violating effects arise from 
the Standard Model, with possible small contributions from new physics. 
Consequently, to establish the existence of new physics effects 
we will need to know the Standard Model effects accurately.  
Fortunately, as the discussion that follows shows, we are moving into 
an era of ``precision'' CKM measurements.  This will hopefully give 
us the elements of the CKM matrix with sufficient accuracy that, 
together with improved theoretical calculations of hadronic matrix 
elements, we will be able to pin down the Standard Model contributions 
and establish the presence of possible new physics.

\section{The Large CKM Matrix Elements}

     The other members of the panel discussion on the CKM matrix, 
M. Artuso{\,}\cite{Artuso00}, P. Faccioli{\,}\cite{Faccioli00}, 
J. Rosner{\,}\cite{Rosner00}, and A. Stocchi{\,}\cite{Stocchi00}, 
have concentrated their contributions on analyses of the small CKM 
matrix elements involving the third generation b-quark and t-quark or 
to making overall fits to the whole matrix. In this section I discuss 
some of the developments involving the CKM matrix elements connecting 
the first and second generation quarks.  These follow closely the 
review by K. Kleinknecht, B. Renk, and myself in the Review of 
Particle Physics{\,}\cite{GKR00}. Detailed references 
can be found there.

\begin{itemize}

\item{The element $|V_{ud}|$ has been most accurately determined 
through analysis of nuclear beta decays that involve transitions 
between states with zero spin for which only the weak vector current 
contributes.  Taking account of higher order radiative corrections 
is essential, and the remaining debate centers on these corrections 
and on whether there is a change in charge-symmetry violation for 
quarks inside nuclear matter at the tenths of a percent level. 
Taking both these uncertainties, a value of 
$|V_{ud}| =  0.9740 \pm 0.0010$ is quoted{\,}\cite{GKR00}.
While the above has been the standard method to obtain $|V_{ud}|$ 
for years, recently there has been an improvement in 
precision of the value obtained from neutron decays.  This has 
fewer theoretical uncertainties, but relies on both the value of 
$g_A /g_V$ and on the neutron lifetime.  Experimental progress 
has been made on the former quantity using very highly polarized 
cold neutrons together with improved detectors.  This results in 
$|V_{ud}| = 0.9728 \pm 0.0012$ from neutron decay, and averaging 
the two independent results for $|V_{ud}|$ gives the value 
$|V_{ud}| = 0.9735 \pm 0.0008$ quoted in Ref.~\cite{GKR00}. This is 
about two sigma lower than expected from unitarity of the first 
row of the CKM matrix, and therefore bears watching.}
  
\item{The matrix element $|V_{us}|$, the sine of the Cabibbo angle, 
is best determined from analysis of $K_{e3}$ decays, which yield 
the value $|V_{us}|  =  0.2196  \pm  0.0023$ ~. Analysis of hyperon 
decays has larger theoretical uncertainties and gives a result for 
$|V_{us}|$ that is not inconsistent. Given the progress in experimental 
techniques and our interest in a more precise check of unitarity, 
a more modern experiment and analysis would be quite worthwhile.}  

\item{In principle, one could determine $|V_{cd}|$ from charm 
decays to non-strange particles, but we lack both high statistics 
data and accurate theoretical input on the relevant form factors.  
The most accurate value presently comes from neutrino and antineutrino 
production of charm off valence $d$ quarks. This yields $|V_{cd}| 
=  0.224 \pm  0.016$~.}

\item{Values of $|V_{cs}|$ can be obtained from neutrino 
production of charm, but they are dependent on assumptions 
about the strange-quark density in the parton sea.  More accurate 
values can be obtained from charm decays to strange particles, 
and in particular $D \to \bar{K} e^+ \nu_e$.  Here the primary 
source of error is in the theoretical estimation of the associated 
form factor and leads to the value $|V_{cs}| = 1.04 \pm 0.16$ ~.  
Significant progress here has recently come from the high 
energy regime at LEP, where direct measurements{\,}\cite{Wcs} of 
$|V_{cs}|$ in charm-tagged $W$ decays give $|V_{cs}| = 0.97 \pm 0.09 
{\rm ~(stat.)} \pm 0.07 {\rm ~(syst.)}$.  This new technique 
already gives a value with a comparable error bar to that from D 
decays.  Furthermore, the $W$ decays into all possible pairs of 
first and second generation quark-antiquark pairs, weighted by the 
squares of the relevant CKM matrix elements.  The result\cite{Sbarra} 
from LEP is that $\Sigma_{i,j} |V_{ij}|^2 = 2.032 \pm 0.032$, where 
the sum extends over $i = u, c$ and $j = d, s, b$.  Since five of 
the six CKM matrix elements involved are well measured or contribute 
negligibly to the sum of the squares, this measurement can also be used 
to obtain a precision measurement of $|V_{cs}| = 0.9891 \pm 0.016$~.}  
The error bar has been reduced by an order of magnitude from that 
obtained using charm decays! 

\end{itemize}
 
\pagebreak

\section{Overall Determination of the CKM Matrix}

     Looking back at the present limits on the accuracy in 
determining the large CKM matrix elements, we quickly see that it is 
primarily not a question of experimental statistics.  Rather, it is 
the systematics and especially the ``theoretical systematics'' that 
limit us.  This becomes even more obvious when we consider the 
determination of $|V_{ub}|, |V_{cb}|$, and $|V_{td}|$.  Time and again 
we need theoretical calculations of matrix elements or parameters that 
relate weak amplitudes at the quark level to those at the hadron level, 
whether it be for inclusive or exclusive decay modes.  

     Theoretical errors in such cases are hard to estimate. It is 
one thing to vary the parameters that enter a given calculation 
over a reasonable range and thereby deduce how the final result will 
vary within a given model.  This is usually easy to do and is a 
minimal estimate of the potential error.  It is quite another 
matter to estimate the effects of what has been left out of the 
model or theory, or to know the accuracy of an {\em ab initio} 
assumption such as quark-hadron duality in a new situation.  

     Furthermore, such errors are generally not Gaussian. 
It still makes sense to quote a reasonable range for a given 
CKM matrix element that corresponds roughly to ``$1~\sigma$'' or 
to ``$90\%$ confidence level,'' but combining several such measurements 
should be done with great care.   As was pointed out during the panel 
discussion, this can have profound physics consequences in making 
an overall fit to the CKM matrix, as those with ``small'' errors 
would say that we are already forced to a non-trivial unitarity 
triangle of ${V_{ub}}^*$, $V_{td}$, and $\sin\theta_{12} V_{cb}$ 
in the complex plane (and therefore a non-trivial phase in the 
CKM matrix and CP-violation in the Standard Model) just from knowledge 
of the lengths of the triangle's sides.  {\em Caveat emptor}. \\
~\\

\noindent{\bf ACKNOWLEDGMENT} \\
~\\
I thank the organizers of Beauty 2000 for their excellent arrangements 
for the meeting and enhancing the interplay of theory and experiment 
at an opportune time. This research work is supported in part by the 
U.S. Department of Energy under Grant No. DE-FG02-91ER40682.


\begin{thebibliography} {99}

\bibitem{KobayashiMaskawa73}M. Kobayashi and T. Maskawa,
	{\em Prog. Theo. Phys.} {\bf 49}, 652 (1973).

\bibitem{Cabibbo63}N. Cabibbo, 
      {\em Phys. Rev. Lett.} {\bf 10}, 531 (1963).

\bibitem{Artuso00}M. Artuso, these proceedings.

\bibitem{Faccioli00}P. Faccioli, these proceedings.

\bibitem{Rosner00}J. Rosner, these proceedings.

\bibitem{Stocchi00}A. Stocchi, these proceedings.

\bibitem{GKR00}F. J. Gilman, K. Kleinknecht, and B. Renk, Review of 
the CKM Quark Mixing Matrix in the 2000 Review of Particle Physics 
of D. E. Groom {\em et al.}, {\em Eur. J. Phys.} {\bf C15}, 110-114 (2000).

\bibitem{Wcs}P. Abreu {\em et al.}, 
      {\em Phys. Lett.} {\bf B439}, 209 (1998); 
      R. Barate {\em et al.}, {\em Phys. Lett.} {\bf B465}, 349 (1999).

\bibitem{Sbarra}C. Sbarra, talk at the Rencontre de Moriond, 
      Les Arcs, France, March 11 - 18, 2000.

\end{thebibliography}
\end{document}